\documentclass[a4paper,12pt]{article}

\title{Stability of S-brane singular solutions and expansion of the universe\break 
}
\author{Riuji Mochizuki
 and Kenji Ikegami
 \\  {\small  Laboratory of Physics, Tokyo Dental College, Chiba 261-8502, Japan}}

\date{\today}

\begin{document}

\maketitle
\abstract{
We investigate stability of single S-brane singular solutions obtained in our previous papers. A stable perturbative solution exists for each of them, while an unstable one exists only if the dilaton field does not depend on time. We apply these perturbative solutions to inflation and late-time acceleration of expansion of the universe. 
\\

\\
keywords: spacelike brane, dilaton expectation value, inflation, 
late-time acceleration
 }
\section{Introduction}

We know that our universe had experienced an exponential expansion during the early time \cite{inflation}. Moreover, the present universe is also expanding exponentially \cite{acceleration}\cite{acceleration2}. 
It is 
desirable that these expansions are explained by use of fundamental theories of elementary particle physics.
Candidates for such theories are ten-dimensional superstring theories and  eleven-dimensional M-theory.  There is, however, a no-go theorem which states that a four-dimensional de Sitter universe can not be realized by the ordinary compactification methods which are applicable to M-theory/superstring theory \cite{no-go1}\cite{no-go2}. One way to overcome this problem would be to put S-branes, which are time-dependent spacelike branes, into the model \cite{sbrane}\cite{sbrane2}\cite{ngreview}. 
Ohta et al. discussed models with S-branes and found that exponential expansions of the universe appeared in some of them.  
Fine-tuned initial conditions were, however, required in order to obtain the enough e-foldings of inflation \cite{fourth-order1}\cite{fourth-order2}.

In our previous papers, we obtained S-brane solutions with a static internal space and  an exponentially expanding external space where the brane exists\cite{mochi}\cite{mochi2}. These solutions are singular solutions, because they can not be obtained by any choice of integral constants in S-brane general solutions \cite{sbgeneral}.

In this paper, we examine the simplest solutions with a static internal space among solutions obtained in our previous papers, namely, $SM_2$, $SM_5$ and $SD_3$-brane solutions, and search for perturbative solutions around them in order to investigate stability of the solutions.  We obtain both  stable  and unstable perturbative solutions. 
Both exist if a dilaton field has constant expectation value, while only stable ones exist if the dilaton field has time-dependent expectation value.
We apply these solutions to inflation of the early universe. 
According to our scenario, we obtain enough e-folding of the universe for successful inflation without fine-tuning of the initial condition. 
Moreover, we apply these solutions to late-time acceleration of expansion of the universe and show that the late-time acceleration lasts for about $4\times 10^{12}$ years in our models.

This paper is organized as follows. In section two, we look for perturbative solutions around the solution obtained in our previous papers. In section three, we apply these perturbative solutions to inflation and late-time acceleration of expansion of the universe.

\hfill\break
\section{Stability of singular S-brane solutions}

Firstly, we present a single singular $S_p$-brane solution with zero dilaton coupling ($SM_p$ and $SD_{p=3}$-brane solutions) according to our previous 
papers \cite{mochi}\cite{mochi2}.  
We consider $D$ dimensional Einstein gravity coupled to a dilaton field $\phi$ and  an 
n-form field strength, $F_{n}$, as a low-energy effective theory of M-theory/superstring theory, 
whose action $I$ is
\begin{equation}
I={c^3\over 16\pi G}\int d^Dx\surd\overline{- g}\Big[ R-{1\over 2}g^{\mu\nu}\partial_{\mu} \phi\partial_{\nu}\phi
-{1\over 2\cdot n!}F_{n}^{\ 2}\Big], \label{eq:action}
\end{equation}
where $G$ is the $D$-dimensional gravitational constant. We assume the following metric form:
\begin{equation}
ds^2=-c^2dt^2+\{a(t)\}^2\sum_{i=1}^{p+1}dx^{i}dx^{i}
+\{c\ b(t)\}^2\sum_{a,b=p+2}^{D-1}j_{ab}d\theta^{a}d\theta^{b},
\end{equation}
where $x^i\ (i=1,\cdots,p+1)$ is a coordinate of the external space where a brane exists, $j_{ab}\ (a,b=p+2,\cdots,D-1) $ is a normalized metric on the compact hyperbolic manifold (CHM)
\cite{hyperbolic1}\cite{hyperbolic2}\cite{hyperbolic3}, which we employ as the internal space, and $\theta^a$ is a coordinate on the CHM.
The field strength for an electrically charged $S_p$-brane is given by
\begin{equation}
(F_n)_{i_{1}\cdots i_{n-1}t}(t)=\epsilon_{i_1\cdots i_{n-1}}\dot{ E}(t),
\end{equation}
where the dot means the time derivative and
\begin{equation}
n=p+2.
\end{equation}
The magnetically charged case is given by
\begin{equation}
(F_n)^{a_1\cdots a_{n}}=\frac{1}{\surd\overline{-g}}\epsilon^{a_1\cdots a_{n}t}\dot{ E}(t),
\end{equation}
where
\begin{equation}
n=D-p-2.
\end{equation}
Equations of motion are
\begin{eqnarray}
%Rij/a^2
{\ddot{ a}\over a}+p\left({\dot{ a}\over a}\right)^2
+(D-p-2){\dot{ a}\over a} {\dot{ b}\over b}
 =
-{(D-p-3)\over 2(D-2)}a^{-2(p+1)} \dot{ E}^2,\cr
%Rab/b^2
{{\ddot{ b}\over b}}
+(D-p-3)\{ \left({\dot{ b}\over b}\right)^2+{K\over b^2}\}
+ (p+1){{\dot{ a}\over a}} {{\dot{ b}\over b}} 
 ={p+1 \over 2(D-2)}a^{-2(p+1)}
 \dot{ E}^2 ,\cr
%-Rtt
(p+1){\ddot{ a}\over a}+
(D-p-2){{\ddot{ b}\over b}}
 =
         -{ D-p-3 \over 2(D-2)}a^{-2(p+1)}
 \dot{ E}^2
       -{1\over 2}\dot{\phi}^2,\cr
%Form 
{d\over d t} \left( a^{-(p+1)} b^{(D-p-2)}\dot{ E}\right)=0,
\qquad\qquad\qquad\cr
%scalar
{d\over d t}(a^{(p+1)}b^{(D-p-2)}\dot{\phi} )=0,
\qquad\qquad\qquad
\label{eq:motion}
\end{eqnarray}
where $K=-1$, which means the internal space is hyperbolic.  
In the case where a condition 
\begin{equation}
\frac{(D-p-3)(p+1)}{2(D-2)}=1 \label{condition}
\end{equation}
is satisfied, the scale factor of the internal space is a constant \cite{mochi}\cite{mochi2}. 
The simplest models satisfying this condition contain a single $SM_2$, $SM_5$ or $SD_3$-brane. Hereafter we consider them.  
An exact solution satisfying the equations (\ref{eq:motion}) and condition (\ref{condition}) is
\begin{equation}
a(t)=a_0e^{\Lambda t},\quad b(t)=b_0,\quad  \dot{ E}={(D-p-3)i\over b_0}({a(t)})^{p+1},\quad \phi(t)=0,
\label{eq:singular}
\end{equation}
where $\Lambda$ and $a_0$ are arbitrary constants and 
\begin{equation}
 b_0={ D-p-3 \over (p+1)|\Lambda |}.
\end{equation}
We note that the scale $a(t)$ of the exernal space exponentially grows with time if $\Lambda>0$.

To study stability of the exact solution (\ref{eq:singular}), we consider a small perturbation around the exact solution. We introduce the following perturbative forms:
\begin{eqnarray}
a(t)&=&a_0e^{\Lambda t}(1+\beta e^{-\alpha t}),
\quad b(t)=b_0(1+{\gamma\over 5}e^{-\alpha t})^{1/2},\cr
\dot{ E}(t)& =& {(D-p-3)i\over b_0(1+\gamma' e^{-\alpha t})^{1/2}}{(a(t))^{p+1}}, \label{eq:perturb}
\end{eqnarray}
where $\beta, \gamma$ and $\gamma'$ are $O(1)$ parameters and $\alpha$ is a constant. 
Here we assume $e^{-\alpha t}<1$. Substituting (\ref{eq:perturb}) to the equations (\ref{eq:motion}) and expanding the equations up to the first order of $e^{-\alpha t}$, we get 
\begin{equation}
\gamma =\gamma' , \qquad 
\dot{\phi}(t)=-(p+1)\Lambda \phi_0e^{-(p+1)\Lambda t},
\label{eq:phi}
\end{equation}
where $\phi_0$ is an arbitrary constant, and
\begin{eqnarray}
&&(\alpha -2(p+1)\Lambda)(\alpha\beta-{\Lambda \gamma\over 2})=0,\cr
&&(\alpha-2(p+1)\Lambda)(\alpha+(p+1)\Lambda)=0,\cr
&&\{ (p+1)(\alpha^2\beta-2\alpha\beta\Lambda+\gamma\Lambda^2)+{\gamma\over 2}\alpha^2\}e^{-\alpha t}\cr
&&\qquad\qquad
 +(p+1)^2\Lambda^2{\phi_0}^2e^{-2(p+1)\Lambda t}
=0. \label{eq:perturbeq}
\end{eqnarray}

If we set $\alpha =2(p+1)\Lambda $ to solve the first and second equations in (\ref{eq:perturbeq}), the third equation gives
\begin{eqnarray}
\gamma =-{(p+1)\over 2p+3}[8p\beta +\phi_0^2],
\end{eqnarray}
which shows that $\phi_0$ is an $O(1)$ parameter. 

On the other hand, if we set $\alpha = -(p+1)\Lambda $ to solve the second equation in (\ref{eq:perturbeq}), we get 
$\beta={\Lambda \gamma\over 2\alpha}=-{\gamma\over 2(p+1)}$
from the first equation and 
$\phi_0=0$ from the third equation.

We have now obtained two perturbative solutions. The first one, which is applicable when $\Lambda t>0$ ($e^{-2(p+1)\Lambda t}<1$), is
\begin{eqnarray}
&&a(t)=a_0e^{\Lambda t}(1+\beta e^{-2(p+1)\Lambda t}),\cr
&& b(t)=b_0(1+{\gamma\over D-p-2}e^{-2(p+1)\Lambda t})^{1/2},\cr
&& \dot{ E}(t) =
 {(D-p-3)i\over b_0(1+\gamma e^{-2(p+1)\Lambda t})^{1/2}}{(a(t))^{p+1}},\cr
&&\dot{\phi}(t)= -(p+1)\Lambda \phi_0e^{-(p+1)\Lambda t},
\label{eq:stable}
\end{eqnarray}
where $\beta, \gamma$ and $\phi_0$ are the parameters satisfying
\begin{eqnarray}
&&\gamma =-{p+1\over 2p+3)}[8p\beta +\phi_0^2].
\end{eqnarray}
The solution (\ref{eq:stable}) approaches the exact solution (\ref{eq:singular}) with $\Lambda t\rightarrow \infty$.
The second one, which is applicable when $\Lambda t'<0$ $( e^{(p+1)\Lambda t'}<1)$, is
\begin{eqnarray}
&&a(t')=a'_0e^{\Lambda t'}(1-{\gamma'\over 2(p+1)} e^{(p+1)\Lambda t'}),\cr
&& b(t')=b_0(1+{\gamma'\over D-p-2}e^{(p+1)\Lambda t'})^{1/2},\cr
&& \dot{ E}(t') =
 {(D-p-3)i\over b_0(1+\gamma' e^{(p+1)\Lambda t'})^{1/2}}{(a(t'))^{p+1}},\cr
&&\dot{ {\phi}}(t')=0,
\label{eq:unstable}
\end{eqnarray}
where $\gamma'$ and $a_0'$ are arbitrary constants.
The solution (\ref{eq:unstable}) approaches the exact solution (\ref{eq:singular}) with $\Lambda t'\rightarrow -\infty$. 

In the next section, we apply the above perturbative solutions  (\ref{eq:stable}) and (\ref{eq:unstable}) to inflation and late-time acceleration of expansion of the universe, so we assume $\Lambda >0$ hereafter. Thus, 
the solution (\ref{eq:stable}) approaches the exact solution (\ref{eq:singular}) with $ t\rightarrow \infty$, so that it expresses stable expansion of the universe.
On the other hand, the solution (\ref{eq:unstable}) approaches the exact solution (\ref{eq:singular}) with $ t\rightarrow -\infty$, so that it  expresses unstable expansion of the universe. It moves away from the exact solution (\ref{eq:singular}) with time, and becomes no use around $t'=0$. 

\section{Application to our universe
}

In this section, we apply the perturbative solutions (\ref{eq:stable}) and (\ref{eq:unstable}) to inflation and late-time acceleration of our universe. Because the $SM_2$ brane extends in 3-dimensional space, which can be regarded to be our universe, we consider the $SM_2$-brane solution only from now on.
As discussed in the previous section, the scalar field (dilaton) is essential in our solutions, but the dilaton field is not contained in the low-energy effective theory of M-theory. 
It is, however, consistent with the action (\ref{eq:action}) containing the dilaton field that the exact solution (\ref{eq:singular}) is the one of the low-energy effective theory of M-theory, as $\phi=0$ in the  exact solution.
On the other hand, the time dependent scalar field $\phi$ in the stable perturbative solution (\ref{eq:stable}) shows that it is not a solution of the effective theory of pure $M$-theory, i.e., the solution with an $SM_2$ brane only, but with both an $SM_2$-brane and  scalar field. We  use the stable solution (\ref{eq:stable}) in order to obtain enough e-folding for successful inflation, so we choose the initial condition of $\phi$ as follows.
To obtain enough e-folding, we need a model in which expansion of the universe lasts a long time during the early time of the universe. 
Even if we set an initial condition $\dot{\phi}=0$,  expansion of the universe might occur according to the stable solution (\ref{eq:stable}). We, however, suppose that the expansion does not last a long time and shift from the stable solution to the unstable one occurs instantly.  Consequently, we adopt $\dot{\phi}\not= 0$ as the initial condition for the scalar field. In this case, expansion of the universe occurs according to the stable solution (\ref{eq:stable}) and lasts a long time, as shown below.

We adopt the following scenario: We choose $\dot{\phi}\not= 0$  as the initial condition for $\phi$, so that the scale $a$ in the metric (2) is increasing exponentially and $\phi$ is decreasing, as shown in the stable solution (\ref{eq:stable}). The unstable perturbative solution (\ref{eq:unstable}) is not applicable to this initial condition. Consequently, the expectation value of $\phi$ must become smaller to be comparable with the quantum fluctuation, i.e., the product of the scalar field $\phi$ and its conjugate momentum becomes comparable with Planck's constant. At this moment, the scalar field $\phi$ can be regarded to be classically equal to zero, and  shift from the stable solution (\ref{eq:stable}) to the unstable solution (\ref{eq:unstable}) occurs due to the quantum effect.  After that, the scale $a$ continues  increasing almost exponentially until the perturbative solution loses its reasonableness. 
In the following, we ignore the expansion of the scale during the shift from the stable solution to the unstable solution, because we expect this period to be much shorter than the time during which the stable or unstable solution continues.

To find out the maximum value of the e-folding, we estimate the rate of the expansion of universe within the above-mentioned period.
We regard the stable perturbative solution (\ref{eq:stable}) to be applicable to the period satisfying $e^{-\Lambda t}<1$ and the unstable perturbative solution (\ref{eq:unstable}) satisfying  $e^{\Lambda t'}<1$. 
So we take the initial time as $t\sim 0$ and the final time as $t'\sim 0$.
If the scalar field is equal to the quantum fluctuation at the time $t=t_1$, $\phi(t_1,x)$ and its canonical momentum $\pi(t_1,x)$ should satisfy the relation
\begin{equation}
\sqrt{-g(t)} \pi (t_1,x) \phi (t_1,x')
\sim   \hbar \delta^{D-1}(x-x'),\label{eq:uncertain}
\end{equation}
which is the so-called Heisenberg-Robertson relation. Here $\hbar\equiv {h\over 2\pi}$, $h$ is  Planck's constant and   
\begin{equation}
\pi(t, x) \equiv {1\over\sqrt{-g(t)}}{\delta I\over \delta \dot{\phi}}
={c^2\over 16\pi G}\dot{\phi}(t,x).
\end{equation}
We integrate (\ref{eq:uncertain}) over the space coordinates $x'$ to obtain
\begin{equation}
V_{D-1}(t_1) \pi (t_1,x) \phi (t_1,x)
\sim  \hbar, \label{uncertain2}
\end{equation}
where
\begin{eqnarray}
V_{D-1}(t_1)&=&\int d^{p+1}xd^{D-p-2}\theta \sqrt{-g(t_1)}\cr
&=&\int d^{p+1}xd^{D-p-2}\theta\ (a(t_1))^{p+1}
(c b(t_1))^{D-p-2}
\end{eqnarray}
is volume of the $(D-1)$ dimensional space at $t=t_1$. 
$\int d^{p+1}x$ is the  $SM_2$ brane world-volume within a radius of particle horizon, 
so that
\begin{eqnarray}
\int d^{p+1}x\sim \left( \int^{t_1}_0 c\ dt'/a(t')\right)^{p+1}.
\end{eqnarray}
In order to calculate the left side of (\ref{uncertain2}) up to the lowest  order of $e^{-2(p+1)\Lambda t}$, we integrate $\dot{\phi}$ in (\ref{eq:stable}) and obtain
\begin{eqnarray}
\phi=\phi_0e^{-(p+1)\Lambda t}.
\end{eqnarray}
Here we require that the fourth equality of (\ref{eq:unstable}) coincides with the fourth equality of the exact solution (\ref{eq:singular}) in the zeroth order of $e^{-2(p+1)\Lambda t}$.  Moreover, we
calculate $V_{D-1}$ up to the lowest order of $e^{-2(p+1)\Lambda t}$ 
\begin{eqnarray}
V_{D-1}(t_1)&&\sim   \ \left( \int^{t_1}_0{c\ dt'\over a(t')}\right)^{p+1} 
(a_0e^{\Lambda t_1})^{p+1}V_{CHM},\cr
&&\sim \ \left( {c\over a_0\Lambda}\right)^{p+1}\left( a_0e^{\Lambda t_1}\right)^{p+1}V_{CHM},
\end{eqnarray}
where $V_{CHM}$ is volume of the CHM.
Substituting these results in (\ref{uncertain2}), we obtain the e-folding until  $t=t_1$
\begin{equation}
e^{\Lambda t_1} \sim \left({(p+1)c^{p+3}\phi_0^2\over 16\pi G_{p+2} \Lambda^{p}\hbar }  \  \right)^{1/(p+1)},\label{eq:efold1}
\end{equation}
where $G_{p+2}\sim {G/V_{CHM}}$ is the $(p+2)$-dimensional gravitational constant in the external space. According to our scenario, the shift to the unstable solution which satisfies $\dot{\phi}=0$ occurs  at $t=t_1$. Because the time during which the shift continues is expected to be so short, we can ignore it, as mentioned above. Considering the condition of the connection between the two perturbative solutions (\ref{eq:stable}) and (\ref{eq:unstable}), we require
\begin{eqnarray}
a_0e^{\Lambda t_1}&=&a'_0e^{\Lambda t'_1} ,\cr
\beta e^{-2(p+1)\Lambda t_1}&=&-{\gamma'\over 2(p+1)}e^{(p+1)\Lambda t'_1},\cr
\gamma &=&\gamma',
\end{eqnarray}
where $t'_1$ is the same time as $t_1$.
These equations lead to
\begin{eqnarray}
\gamma'=-2(p+1)\beta,\quad t'_1=-2t_1,\quad c_1=-3t_1,
\quad a'_0=e^{3\Lambda t_1}a_0 \ .
\end{eqnarray}
Even after the shift, the scale $a$ continues increasing exponentially until 
the unstable solution loses its reasonableness at the time $t'\sim 0$ when $e^{(p+1)\Lambda t'}\sim 1$. After all, we get the total expansion rate of the scale $a$
\begin{equation}
{a'(t'\sim 0)\over a(t\sim 0)}\sim e^{3\Lambda t_1}\sim \left({(p+1)c^{p+3}\phi_0^2\over 16\pi G_{p+2} \Lambda^{p}\hbar }  \  \right)^{3/(p+1)},\label{e-folding}
\end{equation}
where we use (\ref{eq:efold1}). 

Firstly, we apply the above result (\ref{e-folding}) to inflation. Subsitituting $D=11, p=2$ into (\ref{e-folding}) and assuming it to be about $e^{60}$, which is the e-folding of expansion of successful inflation, yield the scale of the internal space
\begin{equation}
c b_0= {2c\over \Lambda }\sim 
 \left( 
{16\pi G_4 \hbar \over 3c^3 \phi_0^2}e^{60}\right)^{1/2}
 \sim 1\times 10^{-21}{\rm m},
\end{equation} 
where we use $\phi_0\sim O(1)$. 
That is to say, we can obtain the enough e-folding for inflation if the scale of the internal space $c b_0$ is about  
$1\times 10^{-21}{\rm m}$. 
Moreover, the duration of the inflation is about
$3t_1\sim {1\over \Lambda}\ln e^{60}\sim 10^{-28}{\rm s}$ in this case. This is within the range  in the literature, for example \cite{Liddle}.

Secondly, we apply the result (\ref{e-folding}) to the late-time acceleration. If dark energy of our universe results from a cosmological constant in 4-dimensional space-time, 
the dark energy density $\rho_{DE}$ and the cosmological constant $\lambda$ have the following relation:
\begin{equation}
I_{\lambda}={c^3\over 16\pi G_4}\int d^4x\{ R+\sqrt{-g}(2-D)\lambda\} =-\int d^4x\{L_G+ \rho_{DE}c^2\}.
\end{equation}
This leads $\lambda={\rho_{DE} 8\pi G\over c^2}$. Moreover, the equation of motion obtained from this action leads to the scale factor
$a(t)\sim $ 
$a_0
\exp({\sqrt{\rho_{DE} 8\pi G_4\over 3c^2}ct})
$. Fitting this to (\ref{eq:stable}) and (\ref{eq:unstable}), we obtain
\begin{equation}
\Lambda \sim \sqrt{\rho_{DE} 8\pi G_4\over 3c^2}.
\end{equation}
Substituting this and the observed density of the dark energy $\rho_{DE}\sim$ \break $ 1\times 10^{-26}{\rm kg/m^3}$ \cite{density} to (\ref{e-folding}), we obtain  
\begin{equation}
3t_1\sim {1\over \Lambda}\ln\left( {3c^5 \phi_0^2\over 16\pi G_4\hbar \Lambda^2}\right)\sim 4\times 10^{12} {\rm yrs}.
\end{equation}
Because the age of our universe is about $1.4\times 10^{10}$ years, this is consistent with the fact that we observe the late-time acceleration at the present time. Moreover, we find that the e-folding of the late-time acceleration of our universe up to the present time is about $2$. This, however, leads to the scale of internal space $cb_0\sim {2c\over \Lambda}\sim 10^{26}{\rm m}$. This value is not  consistent with observation.

\end{document}